\documentclass{sigchi}

\makeatletter
\def\@copyrightspace{\relax}
\makeatother

\usepackage{balance}  

\usepackage{times}

\usepackage{url}      

\makeatletter
\def\url@leostyle{%
  \@ifundefined{selectfont}{\def\UrlFont{\sf}}{\def\UrlFont{\small\bf\ttfamily}}}
\makeatother
\def\pprw{8.5in}
\def\pprh{11in}

\usepackage{booktabs}
\usepackage{paralist}
\usepackage{fixltx2e}
\usepackage{subfigure}
\usepackage{listings}
\usepackage{float}
\usepackage[skip=2pt]{caption}
\usepackage{soul}
\usepackage[normalem]{ulem}

\usepackage{graphicx}
\usepackage{algorithm2e}
\usepackage[noend]{algorithmic}
\usepackage{amsfonts}
\usepackage{amsmath}
\usepackage{ifthen}
\usepackage[usenames]{xcolor, colortbl}
\usepackage{amssymb}
\usepackage{dblfloatfix}
\usepackage{bm}
\usepackage{microtype}

\usepackage{tweaklist}
\renewcommand{\descripthook}{\setlength{\topsep}{0.2\baselineskip}%
  \setlength{\itemsep}{0.1\baselineskip}}
\renewcommand{\itemhook}{\setlength{\topsep}{-0.5\baselineskip}%
  \setlength{\itemsep}{0.1\baselineskip}}

\definecolor{linkColor}{RGB}{6,125,233}
\usepackage[pdftex]{hyperref}
\hypersetup{
pdftitle={Investigating User Interfaces for Quadrotor Camera Motion Control},
pdfauthor={Christoph Gebhardt, Benjamin Hepp, Stefan Stesvic, Otmar Hilliges},
pdfkeywords={robotics; quadrotors; computational design; videography; CHI},
bookmarksnumbered,
pdfstartview={FitH},
colorlinks,
citecolor=black,
filecolor=black,
linkcolor=black,
urlcolor=linkColor,
breaklinks=true,
}

\newcommand{\figref}[1]{Figure~\ref{#1}}
\newcommand{\tabref}[1]{Table~\ref{#1}}

\newcommand{\hidden}[1]{}

\newboolean{showComments}
\setboolean{showComments}{true}
\definecolor{gold}{rgb}{0.80,.60,0}

\newcommand{\todo}[2]{
\ifthenelse{\boolean{showComments}}
{\framebox[\columnwidth][l]{\parbox[l]{3.25in}{\textcolor{red}{#1: #2}}}}
{}
}

\newboolean{showEdits}
\setboolean{showEdits}{false}
\definecolor{amber}{rgb}{1.0, 0.49, 0.0}
\definecolor{bittersweet}{rgb}{1.0, 0.44, 0.37}

\newcommand{\editsCR}[2]{%
\ifthenelse{\boolean{showEdits}}%
{\textcolor{bittersweet}{#1}}%
{#1}%
}

\newboolean{coloredDiff}
\setboolean{coloredDiff}{false}

\newcommand{\ccDiff}[1]{%
\ifthenelse{\boolean{coloredDiff}}%
{\textcolor{bittersweet}{#1}}%
{#1}%
}

\definecolor{LightCyan}{rgb}{0.88,1,1}
\definecolor{LightGreen}{rgb}{0.77,0.93,0.8}
\definecolor{LightOrange}{rgb}{0.95,.67,0.47}
\usepackage[first=0,last=9]{lcg}

\newcolumntype{g}{>{\columncolor{LightGreen}}l}
\newcolumntype{o}{>{\columncolor{LightOrange}}l}

\usepackage{letltxmacro}
\LetLtxMacro{\originaleqref}{\eqref}
\renewcommand{\eqref}{Eq.~\originaleqref}

\begin{document}

\sloppy

\title{WYFIWYG: Investigating Effective User Support in \\ Aerial Videography}

\teaser{
	\centering{
		\includegraphics[width=\linewidth]{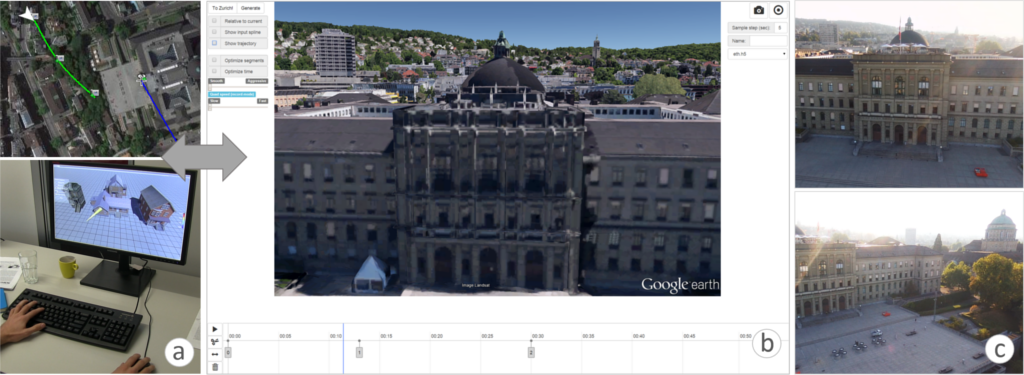}
		\caption{This paper investigates how to effectively support non-expert users in the creation of aerial video shots, comparing (A) the state-of-the-art and (B) WYFIWYG, a tool inspired by expert workflows. (C) The resulting plans can be flown on real robots.}\label{fig:teaser}}
}


\numberofauthors{3}
\author{
	Christoph Gebhardt, Otmar Hilliges\\
	\affaddr{Departement of Computer Science, AIT Lab}\\
	\affaddr{ETH Z\"urich}\\
}
\maketitle


\begin{abstract}
Tools for quadrotor trajectory design have enabled single videographers to create complex aerial video shots that previously required dedicated hardware and several operators. We build on this prior work by studying film-maker's working practices which informed a system design that brings expert workflows closer to end-users.
For this purpose, we propose WYFIWYG, a new quadrotor camera tool which
\begin{inparaenum}[(i)]
\item allows to design a video solely via specifying its frames,
\item encourages the exploration of the scene prior to filming and
\item allows to continuously frame a camera target according to compositional intentions.
\end{inparaenum}
Furthermore, we propose extensions to an existing algorithm, generating more intuitive angular camera motions and producing spatially and temporally smooth trajectories.
Finally, we conduct a user study where we evaluate how end-users work with current videography tools.
We conclude by summarizing the findings of work as implications for the design of UIs and algorithms of quadrotor camera tools.
\end{abstract}



\subsection{ACM Classification Keywords}
I.2.9 Robotics: Autonomous vehicles; Operator interfaces; H.5.2 User Interfaces;


\section{INTRODUCTION}\label{sec:intro}
Cheap and robust quadrotor hardware has recently brought the creation of aerial videography into the reach of end-users. However, creating high-quality video remains a difficult task since users need to control the drone and the camera simultaneously, while considering cinematographic constraints such as target framing and smooth camera motion~\cite{Diaz:2015}.
To automate this difficult control problem, several computational tools for aerial videography have been proposed~\cite{Gebhardt:2016:Airways,Joubert2015,roberts:2016}, casting aerial videography as an optimization problem which takes desired camera positions in space and time as input and generates smooth quadrotor trajectories that respect the physical limits of the robot.
Informed by formative feedback from photographers and filmmakers, this early work focuses on abstracting robot and camera control aspects to be able to plan challenging shots. In this paper we study if and how experts could leverage such tools in their workflows. Based on this formative feedback we design a new system that brings such workflows closer to end-users.

Aiming to translate expert working practices for end-users, we propose WYFIWYG, a new quadrotor camera tool. Based on the findings of formative interviews with film-makers and quadrotor operators, we implemented a UI that
\begin{inparaenum}[(i)]
\item enables users to design a video solely via specifying its frames (hiding quadrotor-related aspects like force diagrams or a 2D-trajectory),
\item a camera control mechanism that encourages the exploration of a scene and
\item a keyframe sampling method allowing to \emph{continuously} frame a camera target according to compositional intentions.
\end{inparaenum}

In addition, we extend an existing algorithm \cite{Gebhardt:2016:Airways} to generate more intuitive angular camera motions and to improve the overall smoothness of quadrotor camera trajectories.
Finally, we conduct a user study in which we evaluate WYFIWYG and a state-of-the-art tool \cite{Joubert2015}. A key-finding is that current tools complicate the design of globally smooth video shots by requiring users to specify keyframes at equidistant points in time and space. We conclude by summarizing implications for UI and optimization scheme design that are important to support users in creating aerial videos.

In summary, we contribute:
\begin{inparaenum}[1)]
  \item An analysis and discussion of formative expert interviews.
  \item A new UI design for aerial videography.
  \item Extensions to an existing quadrotor camera trajectory optimizer \cite{Gebhardt:2016:Airways}.
  \item A discussion of implications for future UI and algorithmic research based on the study results.
\end{inparaenum}

\section{RELATED WORK}\label{sec:rw}
\subsection{Robotic Behavior Control}
Automating the design of robotic systems based on high-level functional specifications is a long-standing goal in graphics and HCI. Focusing on robot behavior only, tangible UIs \cite{Zhao:2009:MCP}, and sketch based interfaces to program robotic systems \cite{Liu:2011:RMS,Sakamoto2009} have been proposed. Recently, several works introduce gestures as a mean for human-drone interaction  \cite{Cauchard:2015,E:2017}.

\subsection{Camera Control in Virtual Environments}
Camera placement \cite{Lino:2012}, path planning~\cite{CAV:CAV398,Li:2008:SG} and automated cinematography~\cite{Lino:2011:Director} have been studied extensively in the context of virtual environments, for a survey see~\cite{Christie2008}. Many of these papers identify the need for suitable UI metaphors so that intelligent cinematography tools can support film makers in the creative process. Most notably the requirement to let users define and control the recorded video as directly as possible, instead of controlling the camera parameters (e.g., ~\cite{Drucker94intelligentcamera,Lino:2015hn,Lino:2011:Director}). In this context it is important to consider that virtual environments are not limited by real-world physics and robot constraints, hence can produce camera trajectories that could not be flown by a quadrotor.

\subsection{Trajectory Generation}
Quadrotor motion plan generation is a well studied problem and various approaches have been proposed, including generation of collision-free plans applied to aerial vehicles~\cite{US:2003vc, Richards:2004kb}, global forward planning approaches to generate minimum snap trajectories \cite{Mellinger2011a}, or real-time methods for the generation of point-to-point trajectories~\cite{Mueller:2013:vm}.

\subsection{Computational Support of Aerial Videography}
With the increasing popularity of aerial videography a number of tools to support this task exist. Commercial applications are often limited to placing waypoints on a 2D map~\cite{apm:2015,dji:2015:groundstation,litchi:2016}. 

Several algorithms for the planning of quadcopter trajectories, taking both aesthetic objectives and the physical limits of the robot into consideration, have been proposed. These tools allow for the planning of camera shots in 3D~\cite{Gebhardt:2016:Airways,Joubert2015,roberts:2016}.
\begin{figure}[tbh]
	\centering
	\includegraphics[width=0.9\linewidth]{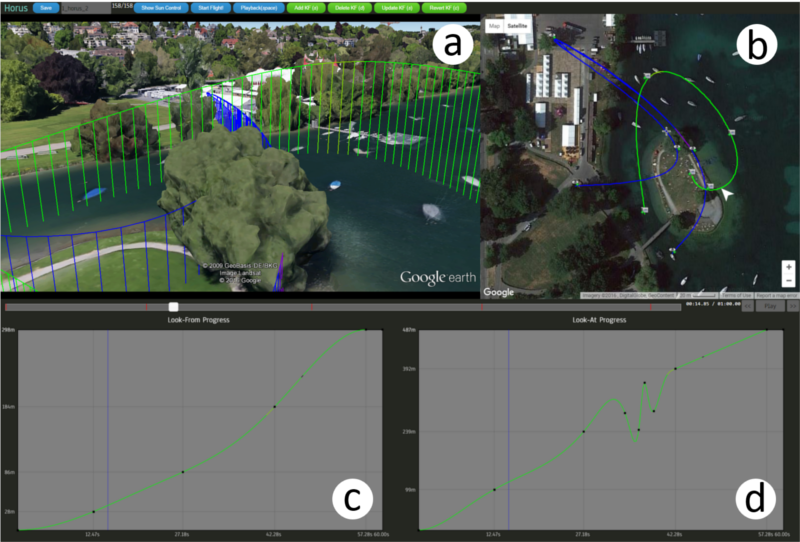}
	\caption{Horus \protect\cite{Joubert2015}
		visualizes a user-specified trajectory in 3D (a) and 2D (b). Two plots visualize progress over time for the look-from / quadrotor (c) and look-at / camera targets (d) trajectories and allow users to change timing of a video.}
	\label{fig:horus}
\end{figure}
Airways \cite{Gebhardt:2016:Airways} allows users to specify keyframe-based trajectories and select a camera target for each keyframe. 
After generation users can inspect the trajectory and see a video preview. 
With Horus \cite{Joubert2015} users can specify a camera trajectory using a 3D preview or a 2D map (see \figref{fig:horus}). The tool offers progress curves for quadrotor and camera target positions, allowing users to change the timing of a video.
Horus can detect but not correct violations of the limits of the robot model. 
In contrast, \cite{roberts:2016} proposes a method which takes physically infeasible camera paths as input and generates quadrotor trajectories that match the intended camera motion as closely as possible.

\cite{Joubert2015} conducted an evaluation of their tool with cinematographers. We study aspects pertaining to end-users and contribute new insights on quadrotor videography form this perspective.

Recently, several works have been published which cover the generation of quadrotor camera trajectories in real-time to record dynamic scenes. Real-time performance is attained by planning only locally \cite{Naegeli2017, Nageli:2017:SIGGRAPH} or by reducing the problem to a lower-dimensional subspace \cite{galvane2016automated, joubert2016towards}. In contrast to these papers, our work focuses on the generation of quadrotor motion for city or landscape shots.



\begin{figure*}[t]
	\centering
	\includegraphics[width=1.0\linewidth]{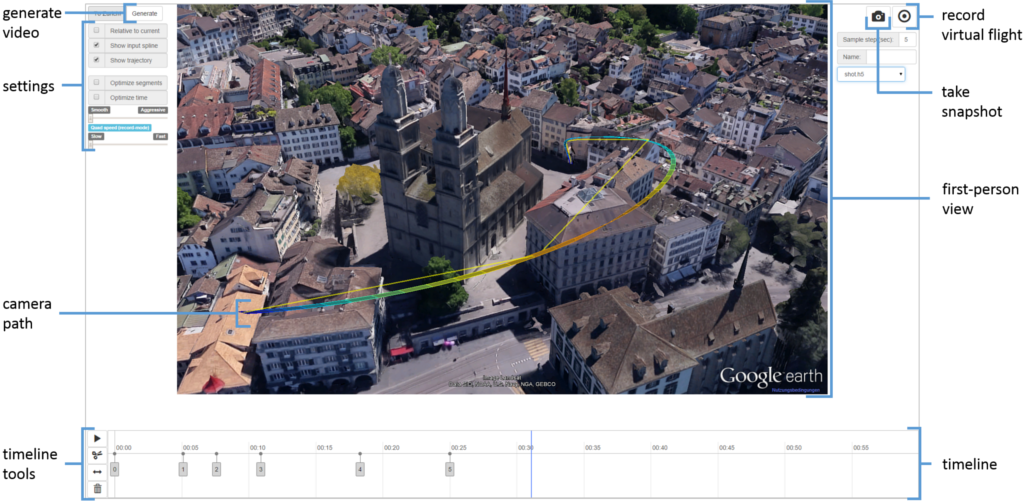}
	\caption{In WYFIWYG users can define keyframes in first-person view. They can add keyframes to a video by taking a snapshot of the current view or recording a virtual flight. A timeline enables the adjustment of a shot's timing.}
	\label{fig:videoui}
\end{figure*}

\section{FORMATIVE INTERVIEWS}\label{sec:context_analysis}
To inform our design, we conducted a series of expert interviews. Here we report on aspects which experts defined as being crucial for creating pleasing aerial video and which are not, to a satisfying extend, supported in existing tools.

We interviewed six professional users including three aerial videographers, producing for instance footage for real estate agencies and other commercial purposes, two professional camera men working on TV, movie and documentary sets and one quadrotor operator specialized on high-quality commercials and Hollywood film productions.
We visited our participants in their offices or workshops during their working hours to understand their workflows, workplaces and the equipment and tools used for the planning and the execution of aerial video shots.
The interviews were not restricted in duration and typically lasted between 1 and 2 hours. The interviews were semi-structured around questions on planning procedures, workflow and tool use. In addition, we introduced the participants to two existing quadrotor camera tools \cite{Gebhardt:2016:Airways,Joubert2015} via the original videos. We then asked the experts to explore with us if and how these tool could support existing workflows and which additional features would be desirable.
While our experts also stated aspects already mentioned in literature \cite{Joubert2015}, we now highlight previously unreported results.

\subsection{Target Framing}
The ability to control and fine-tune the framing of a filmed subject \emph{continuously} and with high-precision is an essential aesthetic tool. The interviewees highlighted the importance of being able to precisely position an object in the image plane subject to a compositional intention (e.g., a simultaneously moving foreground and background).
For this reason, aerial video shots are usually taken by two operators, one piloting the quadrotor and one controlling the camera, allowing to constantly fine-tune the subject framing.
Several professional operators also stated that following a specific quadrotor trajectory is not a primary concern, or in the words of one of our participants \emph{``what counts is the result [video], not the trajectory of the quadrotor''}. For instance, even when circling a filmed object, one participant explained that this is always performed based on the live camera stream and flying a perfect circle may even be counterproductive.

\subsection{Smooth Camera Motion}
The key to aesthetically pleasing aerial video is described by one of our participants as \emph{"[...] the camera is always in motion and movements are smooth"}. Another expert stated that smoothness is considered \textit{the} criteria for shots with a moving camera (see also \cite{Audronis:2014,Hennessy:2015}), whereas the dynamics of camera motion should stay adjustable. We stress this point since current algorithms keep the temporal position of keyframes fixed, hence can only generate smooth motion locally and produce varying camera velocities in-between different sections of a trajectory (see section Method, Smooth Camera Motion).          

\subsection{Exploration}
In practice, aerial shots are often defined in-situ in an exploratory fashion. In professional settings so-called `layout-drones' are used to initially record a scene from various perspectives and only after reviewing the results, high-end equipment is used for the final shot.
Most interviewees stressed that this phase is of fundamental importance to find good shots.




\section{USER INTERFACE DESIGN}\label{sec:user_interface}
Based on above findings, we propose a new tool, aiming to translate expert working practices for end-users via an easy-to-use UI design. In the following, we will explain UI, camera control, and virtual flight mode of WYFIWYG and highlight how they are derived from the expert interviews.

\subsection{Video UI}
To reduce complexity we design the UI in a way that it transforms the general task of specifying a robot movement plan into a task more akin to creating a video. Therefore, we take the design decision to hide all quadrotor-related aspects like a 2D-trajectory or input-force diagrams.
Users see the virtual world through a first-person-view and can freely position this view within a 3D virtual environment (see \figref{fig:videoui}). Once satisfied with a viewpoint, it can be added to the timeline as a video frame.
After each keyframe insertion, an optimization algorithm generates a trajectory and the resulting video can be previewed immediately. 
Similar to common video editing tools, we also provide a timeline and functionality to edit the shot timings (e.g., moving keyframes in time). 
Due to this example-centric approach our tool does not provide an editable trajectory visualization (camera path is still rendered in 3D) and users need to specify keyframes in the image plane to design a video.
Taking up the "circling around an object" example from the expert interviews, we designed our UI to lead users in positioning keyframes based on what they see in the preview, focusing on framing and not worrying about the geometric shape of the trajectory.

\subsection{Integrated Camera Control}\label{sec:camera_control}
Unpacking the need for precise target framing, experts highlighted that in professional settings, two operators work together to adjust a camera's position as well as its pitch and yaw angle simultaneously. To enable a similar way of working in our single-user tool, we provide a control mechanism which integrates translational and rotational degrees of freedom. Research has shown that integrating translational and rotational degrees of freedom gives users more fine-grained control over 3D movements \cite{Zhai1998} and should lead to better compositional abilities when framing a camera target. For our tool, we implemented a 3D-camera control which can be used with a variety of input devices that allow for simultaneous control of 5-DoF (quadrotor cameras do not allow for roll), such as game pads or multi-touch controls (cf. video). 
In addition, the experts also highlighted the importance of environment exploration for finding interesting perspectives and planning an aesthetically pleasing camera path. By providing an integrated camera control in combination with a first person view, users can virtually fly through the 3D scene like in a flight simulator. With this gamified interaction, we intent to encourage users to explore the environment when designing a shot.
In contrast, Airways only shows a 3D preview after trajectory generation. Horus offers a preview at planning time which would generally allow for exploration. Nevertheless, we believe that mouse interaction (which separates translational and rotational movement) makes exploration cumbersome compared to a gamepad.


\subsection{Virtual Flight}\label{sec:virtual_flight}
A final finding relates to the need to continuously re-fine subject framing over an entire shot. To allow for continuous target framing, we implemented an extension to the basic keyframe-based setting which we dub \emph{virtual flight} mode. In this mode, the user directly records the entire shot by flying in first person view through the virtual environment (without specifying discrete keyframes).
Behind the scenes, we automatically sample the camera's position and orientation (at an adjustable time interval).
Our algorithm adopts the positions of the virtual camera motion, optimizing and smoothing only its dynamics. Based on the suggestion of a participant, the resulting motion plan can also be played-back and edited in situ to fine-tune target framing.
This mode lends the paper its title: WYFIWYG or "what you fly is what you get".



\section{METHOD}\label{sec:method}
In addition to the UI design we also contribute extensions to existing trajectory generation methods allowing for more fine-grained target framing and easy creation of smooth camera motion. Our algorithm is based on the method presented in \cite{Gebhardt:2016:Airways}. A recap can be found in this paper's appendix.

\subsection{Target Framing}
The context analysis highlighted the importance of fine tuning target framing. In the real world setting the camera is oriented and positioned to align a target in image plane, in order to achieve a desired compositional effect. In contrast, Airways and Horus orient the camera based on user-defined target positions and generate a look-at trajectory in-between them. In Airways, these look-at positions are always centered in image plane, taking away all compositional abilities. Horus provides the possibility to adjust target framing by moving a camera's look-at position with respect to a camera target. Nevertheless, orienting the camera based on a look-at trajectory can yield undesirable effects. First, optimizing the camera orientation based on a shortest path interpolation in-between look-at positions can cause unexpected camera tilting.
\begin{figure}[tbh]
	\centering
	\includegraphics[width=1.0\linewidth]{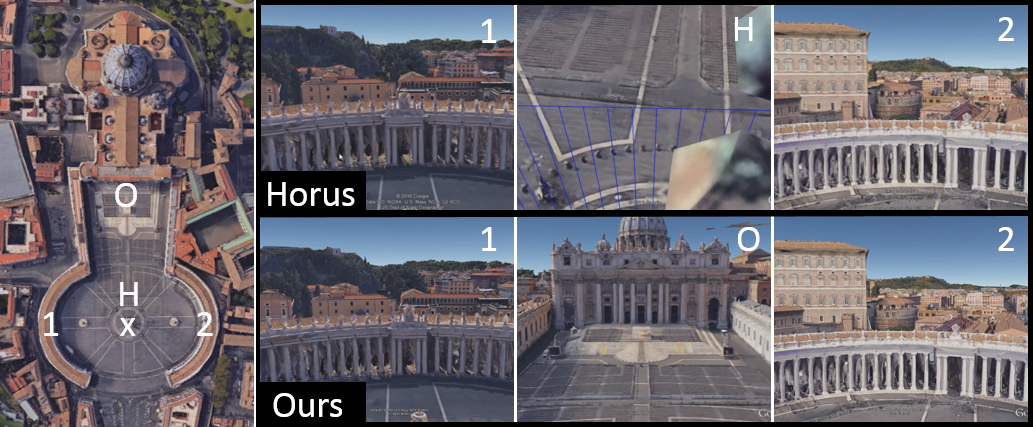}
	\caption{On the left the position of the virtual camera (x), the two specified look-at positions (1, 2) and the look-at position of  the generated intermediate frames (H,O) are shown. The first row on the right shows the generated video of Horus with a camera tilt due to the shortest path interpolation between the two look-at points (H). The second row shows the result of our optimization method for the same input, framing St Peter's Basilica in the middle of the shot (O, cf. video).}
	\label{fig:target_framing}
\end{figure}
We illustrated this problem in \figref{fig:target_framing}, where the shortest path interpolation in-between keyframes causes the camera to miss large parts of St Peter's Basilica\footnote{The example is chosen specifically to visualize the problem.}. A problem which occurs more often are undesirable camera dynamics. Orienting the camera based on a timed trajectory causes its motion to be faster when the reference point on the trajectory is close to the position of the camera and slower when the reference point is more distant. Although, in both cases the covered distance in camera angle is the same, thus smooth camera motion could be generated\footnote{See video from 1:50 min to 2:50 min.}.
To overcome these problems, we model pitch and yaw angle of the camera (roll is not desired in a videography setting) and optimize them based on the orientation of the virtual camera of user-specified keyframes. Modeling the gimbal with:
\begin{gather}
	\label{eq:gimbal_model}
	\dot{\psi_{\mathit{g}}} = u_{\mathit{g},\psi} \\
	\dot{\phi_{\mathit{g}}} = u_{\mathit{g},\phi} \nonumber \\
	[\psi_{\mathit{g},\mathit{min}}, \phi_{\mathit{g},\mathit{min}}]^{T}
	\leq [\psi_{\mathit{g}}, \phi_{\mathit{g}}]^{T}
	\leq [\psi_{\mathit{g},\mathit{max}}, \phi_{\mathit{g},\mathit{max}}]^{T} \label{eq:gimbal_bounds} \\
	\mathbf{u}_{\mathit{g},\mathit{min}}
	\leq [u_{\mathit{g},\psi}, u_{\mathit{g},\phi}]^{T}
	\leq \mathbf{u}_{\mathit{g},\mathit{max}} , \nonumber
\end{gather}
where the inputs $u_{\textit{g},\psi}$, $u_{\textit{g},\phi}$ represent the angular velocities
of the yaw $\psi_{\mathit{g}}$ and pitch $\phi_{\mathit{g}}$ of the gimbal
and both the inputs and the absolute angles are bounded according to the dynamics and range-of-motion of the physical gimbal.
Using this gimbal model, we now introduce an additional cost-term
\begin{equation}
\label{eq:orientation_cost}
E^{o} = \sum_{j=1}^{M} || (\psi_{\mathit{g},\eta(j)} + \psi_{q,\eta(j)}) - \psi_{j} || ^ {2} + \sum_{j=1}^{M} || \phi_{\mathit{g},\eta(j)} - \phi_{j} || ^ {2} .
\end{equation}
Where $\psi_{j}$ and $\phi_{j}$ are the desired yaw and pitch orientation of the camera at each keyframe, $\psi_{\mathit{g},\eta(j)}$, $\psi_{\mathit{q},\eta(j)}$ and $\phi_{\mathit{g},\eta(j)}$ are the gimbal and quadrotor yaw angle as well as the gimbal pitch angle at a keyframe's corresponding time point on the trajectory. By modeling the yaw angle of the quadrotor and the gimbal separately and adding it up in \eqref{eq:orientation_cost}, the generated trajectories can exploit the full dynamic range of the quadrotor and the gimbal around the world frame z-axis. Furthermore, by separating the reference tracking of pitch and yaw in \eqref{eq:orientation_cost}, we can prevent undesired camera tilt in-between keyframes for most cases (see example in the bottom row of \figref{fig:target_framing}).
We now rewrite the gimbal model \eqref{eq:gimbal_model} as a discretized first-order dynamical system, formulate this system as equality constraints, state its bounds (\eqref{eq:gimbal_bounds}) as inequality constraints and incorporate both into the original optimization problem (\eqref{eq:quadratic_program}, appendix). We add $E^{o}$ to the objective function of \cite{Gebhardt:2016:Airways} and include a penalizing term on higher derivatives of the yaw angles and the gimbal pitch (cf. \eqref{eq:quad_derivative_cost}, appendix).

In the original method the non-linearities introduced by the camera target tracking required the usage of a computationally expensive iterative quadratic programming scheme~\cite{Gebhardt:2016:Airways}. In contrast our method remains quadratic and can be solved directly. This reduces optimization run times for camera target tracking problems from tens of seconds to seconds (a camera trajectory with 20 seconds runtime is generated in 2 seconds compared to 14 seconds with ~\cite{Gebhardt:2016:Airways}).

\begin{figure}[tbh]
	\centering
	\includegraphics[width=\linewidth]{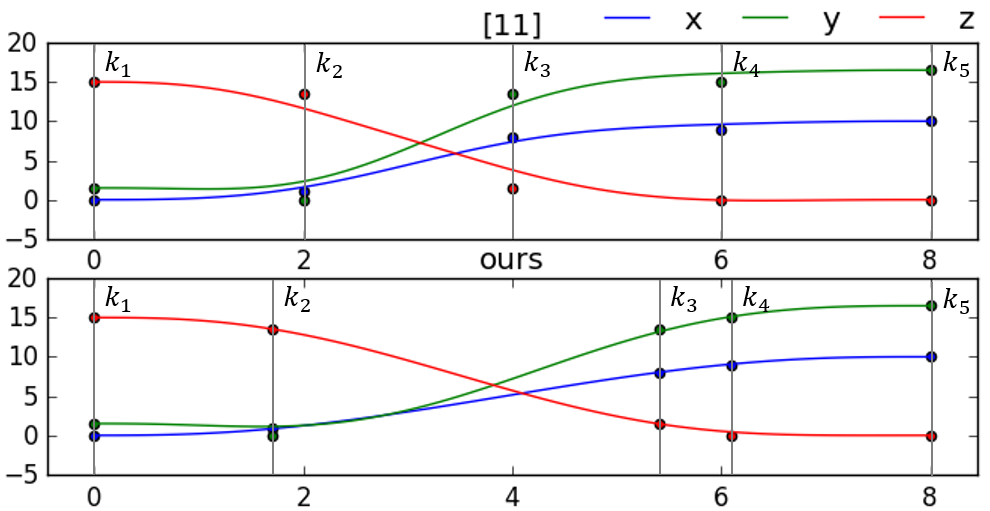}
	\caption{Comparison of trajectory generation methods. Compared to \protect\cite{Gebhardt:2016:Airways}, our method adjusts timings to better fit positional distances of keyframes ($k_1,...,k_5$).}
	\label{fig:time_optimization}
\end{figure}

\begin{figure*}[t]
	\centering
	\includegraphics[width=1.0\linewidth]{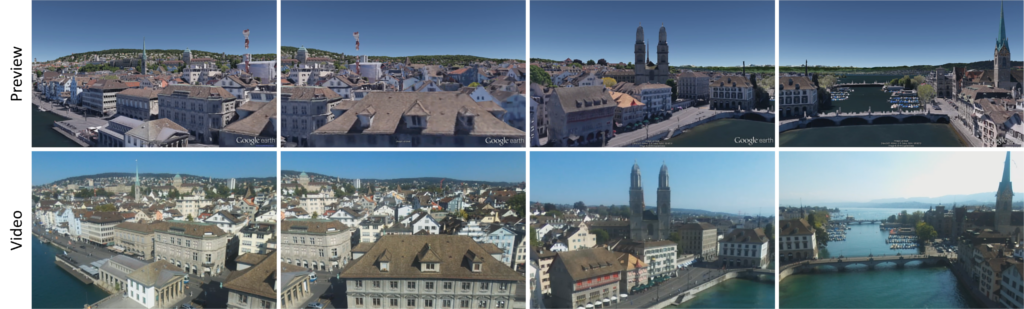}
	\caption{Visual results of our method. Top: snapshots from planning tool. Bottom: corresponding results from real quadrotor.}
	\label{fig:preview_video}
\end{figure*}

\subsection{Smooth Camera Motion}
Smooth camera motion over an entire sequence is a quality criteria for aesthetically pleasing aerial videos (see expert interviews). With current tools' optimization schemes this is not easy to achieve since the timings of user-specified keyframes are kept fix and are not optimized when generating a trajectory. Therefore, the resulting camera motion is only smooth locally and still can vary in-between keyframes which results in visually unpleasant video\footnote{See video from 2:54 min to 3:08 min.}. To generate smooth motion over an entire shot with the existing tools, users need to ensure that the ratio of distance in time to distance in space is similar in-between all keyframes.
\cite{Joubert2015} tackles this problem by providing look-at (camera look-at position) and look-from (quadrotor position) progress curves, allowing to edit the relative progress on a trajectory over time. An even slope over the entire curve indicates a smoothly moving camera. Nevertheless, the effect of manipulating these progress curves on camera motion can sometimes be difficult to understand (see \figref{fig:horus}, d). To help even novice users to produce globally smooth temporal behavior, we extend our method to not only optimize the positions of keyframes in space but also in time. This can be stated as
\begin{align}
	\label{eq:time_optimization}
	\underset{\mathbf{t}}{\text{minimize}} \ & \ f(\mathbf{t}) + Nw \\
	\text{ subject to } \ & \ t_{i-1} < t_{i} < t_{i+1} ,\label{eq:time_constraint}
\end{align}
where $f(\mathbf{t})$ is the minimum of the objective function of \cite{Gebhardt:2016:Airways} for the keyframe times $\mathbf{t} = [t_2, t_3, ..., t_{M}]$ ($t_1$ is always $0$ and not optimized) and $w$ is a user specified weight factor. $N$ is the number of discrete time steps and an implicit decision variable as it depends on the last keyframe time.

Intuitively, setting the weight $w$ allows users to trade-off smooth but long with aggressive but short trajectories (in time). For example, setting $w > max(D^{3} x_{i})$ (the maximum jerk in a single time step), would force the quadrotor to fully exhaust its force limits in each time step. Making $N$ an optimization variable and including weight $w$ for each discretized step prevents degenerate solutions of infinitely long trajectories, where the optimization adds steps with $D^{3} x_{i} \approx 0$ which are free with respect to the optimization's objective.
In case users want to optimize the segment timings of fixed length trajectories, the formulation also allows to remove the last keyframe $t_m$ from \eqref{eq:time_optimization} and set $w$ to zero (following \cite{Mellinger2011a}).
\eqref{eq:time_optimization} is solved via gradient descent. The directional derivatives for each keyframe denoted by $g_i$ are computed numerically
\begin{equation}
\nabla_{g_i} f = \frac{f(\mathbf{t} + hg_i) - f(\mathbf{t})}{h} ,\nonumber
\end{equation}
where $h$ is a small number and $g_i$ is constructed in such a way that the $i$th element is $1$ and all other elements are $0$. By summing up the directional derivatives $\nabla_{g_i} f$ of all keyframes we compute the gradient $\nabla_{g} f$.
We then perform gradient descent via line-search on the optimization problem of \eqref{eq:time_optimization}, enforcing its constraint \eqref{eq:time_constraint}.
\figref{fig:time_optimization} illustrates the effect of this time optimization by comparing our approach with the standard method.
For the same set of keyframes and optimization weights as well as a fixed trajectory end time, our method adjusts the timings such that larger positional distances in-between keyframes are reflected by larger temporal distances. This leads to a better positional fit with the reference $x, y, z$-coordinates of the keyframes (e.g., see z-coordinate of $k_3$).
To compare smoothness between both methods quantitatively, we calculate the accumulated jerk of both trajectories normalized by the horizon length\footnote{Minimizing jerk is common practice to smoothen motion (cf. \cite{flash1985coordination})}. This measures is smaller for our method (ours: 1.73$\frac{m}{s^3}$, \cite{Gebhardt:2016:Airways}: 2.63$\frac{m}{s^3}$), indicating a smoother camera motion.
Note that the global time optimization prevents real-time performance. However, it is fast enough to be employed in the user study.

\subsection{Visual Results}
We evaluate the functionality of our system  qualitatively by designing a number of aerial video shots and executed the resulting plans on a real quadcopter (unmodified Parrot Bebop 2). \figref{fig:preview_video} shows selected frames from the preview and resulting footage (cf. accompanying video).


\section{Evaluation}\label{sec:study}
To better understand the effectiveness of particular UI- and optimization scheme features in terms of supporting end-users in the creation of aerial footage, we conduct a preliminary user study where we evaluate two variants of our system and Horus \cite{Joubert2015} (see \figref{fig:horus}). This tool was chosen as representative of the-state-of-the-art, since other work either solely focuses on the optimization aspects of quadrotor camera tools \cite{Gebhardt:2016:Airways} or is not available as open-source \cite{roberts:2016}. 

\emph{Participants:} Twelve participants (5 female, 7 male) were recruited from our institution (students and staff). The average age was 25.3 (SD = 3.1, aged 19 to 32). We included one expert, working part-time as a professional quadrotor operator, the remaining participants reported no considerable experience in aerial nor normal photo- or videography. Five participants reported prior experience with 3D games, four had limited experience and three reported no experience.

\emph{Experimental conditions:} We investigate \textit{Horus} and two variants of WYFIWYG. The first variant takes keyframes from the basic snapshot-mode as input (\textit{snapshot}). In the second variant, users directly specify the camera path (equidistant keyframe sampling) (\textit{virtual-flight}).
\textit{Horus} is controlled via mouse and keyboard, whereas \textit{snapshot} and \textit{virtual-flight} are controlled using a gamepad.
We use a within-subjects design with fully counterbalanced order of presentation to compensate for learning effects.

\emph{Tasks:} The study comprises two tasks:
\begin{inparaenum}[1)]
	\item Participants were asked to faithfully reproduce an aerial video shot shown to them by the experimenter (T1). The shot was designed with the help of an expert as a shot only possible with airborne camera. 
	\item Participants were asked to design a video of their liking with a maximum duration of one minute (T2).
\end{inparaenum}

\emph{Procedure:} In the beginning, participants were introduced to the systems and asked to design a short video in each condition. During this tutorial they could ask the experimenter for help. After that participants first solved T1 and then T2, each in all conditions.
Both tasks were completed when participants reported to be satisfied with the similarity to the reference (T1) or the designed video (T2).
Participants were encouraged to think aloud.
For each task and condition participants completed the NASA-TLX and a questionnaire on satisfaction with the result and the system. 
At the end an exit interview was conducted.
A session took on average 92 min (SD = 29 min) (tutorial $\approx$ 26 min, T1 $\approx$ 29 min, T2 $\approx$ 22 min).

\section{RESULTS}\label{sec:results}
Here we discuss quantitative results of our study (for further results see Appendix B). 
Following \cite{Cumming01012014,dragicevic2016fair}, we abstain from null hypothesis significance testing and report interval estimates\footnote{standard deviation = SD, 95\% confidence interval = CI.}.
We test conditions according to the findings of the expert interviews and analyze their usability and user experience.

\subsection{Target Framing}
\begin{figure}[tbh]
	\centering
	\includegraphics[width=1.0\linewidth]{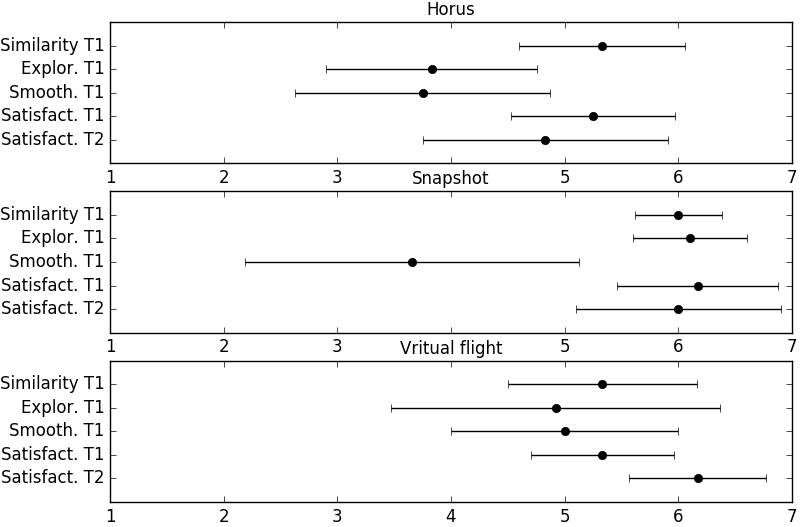}
	\caption{Visualizing participant responses and their CIs.}
	\label{fig:boxplots}
\end{figure}

In T1 we asked participants to reproduce a given video. The idea is that by setting the reference and comparing video similarity, we are able to reveal potential advantages and drawbacks of the different target framing approaches used in our conditions. To quantitatively assess similarity of videos from T1 we compare resulting trajectories with the reference. Due to differences in underlying algorithms, we only compare trajectory positions and not their dynamics. We normalize the length of all trajectories to the duration of the reference. 
\begin{figure}[tbh]
	\centering
	\includegraphics[width=0.9\linewidth]{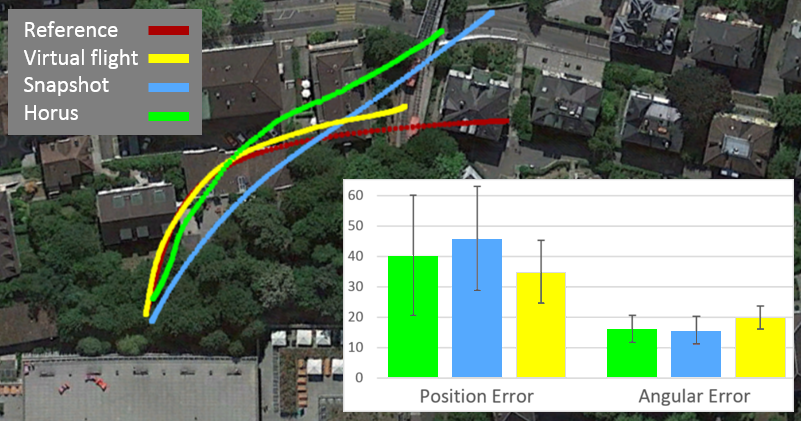}
	\caption{Visualization of the average trajectory of each condition and the reference. Inset shows average errors and CIs.}
	\label{fig:average_trajectories}
\end{figure}
\figref{fig:average_trajectories} plots the average trajectories by UI in comparison to the reference. The inset summarizes position and orientation error over all trajectories and users. Initially, participants perceived \textit{virtual-flight} as difficult to control. However, on average this mode produces the closest positional match with the lowest mean error and the tightest CI. It is followed by \textit{Horus} and \textit{snapshot}. For the angular error \textit{Horus} and \textit{snapshot} have the best result followed by \textit{virtual-flight}.
\figref{fig:boxplots} shows participant responses on perceived similarity to the reference video on a scale from 1 (very different) to 7 (very similar).
Comparing means and confidence intervals in between all conditions for positional and angular error as well as for perceived similarity, no significant quantitative differences in target framing can be determined for the given task. 
Nevertheless, using \textit{Horus} two participants mentioned their struggle with unintended camera tilt and non-smooth camera motion as effects of optimizing target framing based on look-at positions (referring to section Method, Target Framing). Both were not able to generate the video they intended to design.

\subsection{Smooth Camera Motion}
\figref{fig:boxplots} summarizes rankings of the perceived smoothness on a scale ranging from 1 (non-smooth) to 7 (very smooth).
Our participants regularly adjusted the timing of shots to attain smooth camera motion. As expected, several participants (not the expert) had problems to attain globally smooth camera motion paths. They were not able to position keyframes such that the ratio of distance in time to distance in space is similar, resulting in non-smooth footage (see video from 3:54  to 4:08 min). In this context, observations and participants thinking-aloud revealed that most of them expected the optimization to generate smooth camera motion over all specified keyframes. However, only few used the global time optimization, which actually provided this functionality.
This may be due to 
\begin{inparaenum}[(i)]
	\item  the longer runtime of the  procedure and
	\item this being an on-demand feature and participants may not have been aware of it (although shown in the tutorial).
\end{inparaenum}
The two participants that did use the feature were very positive about its utility in particular after discovering that with this method fewer keyframes are necessary to achieve appealing videos. Both used the segment times optimization such that the temporal length of the original and the time-optimized motion path stays the same. Still, jerk and angular jerk of the time-optimized trajectory is smaller in both cases, 
compared to the trajectory generated by using unmodified \cite{Gebhardt:2016:Airways} (see \tabref{tab:jerk}), quantitatively verifying smoother camera motion.
\begin{table}[tbh]
	\centering
	\setlength{\tabcolsep}{3pt}
	\begin{tabular}[c]{|l||l|l|l|}
		\hline
		Participant&Method&Jerk ($\frac{m}{s^3}$)&Angular jerk ($\frac{\circ}{s^3}$)\\ 
		\hline
		1&\cite{Gebhardt:2016:Airways}&0.07& 2.29\\ 
		&time-opt.&0.06&0.04\\ 
        \hline
        2&\cite{Gebhardt:2016:Airways}&1.15&4.01\\ 
		&time-opt.&0.74&3.44\\ 
		\hline
	\end{tabular}
	\caption{Comparison of jerk and angular jerk for trajectories generated with \protect\cite{Gebhardt:2016:Airways} and with our time optimization.}
	\label{tab:jerk}
\end{table}

\subsection{Exploration}
\begin{figure}[tbh]
	\centering
	\includegraphics[width=1.0\linewidth]{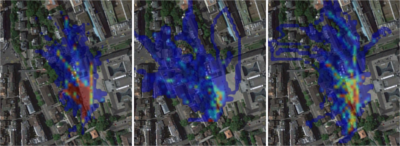}
	\caption{Heatmap of camera positions in \textit{Horus} (left), \textit{snapshot} (middle) and \textit{virtual-flight} (right).}
	\label{fig:heatmaps}
\end{figure}

To assess support for freeform exploration, we logged the camera positions over all participants in T1. This is visualized as heatmap in \figref{fig:heatmaps}, clearly showing that participants cover more ground and experiment more in both WYFIWYG conditions than with Horus.
This is also reflected in the participants perception. On a scale from 1 (does not encourage exploration) to 7 (strongly encourages exploration), they rated \textit{snapshot} first, followed by \textit{virtual-flight} and \textit{Horus} (cf. \figref{fig:boxplots}).
Users commented that being able to evaluate different perspectives quickly helped in solving T1 as they could better match which views were present in the reference.

\subsection{Usability}
To asses usability differences between the three tools we asked our participants to fill out the NASA-TLX questionnaire. Looking at the NASA-TLX scores, summarized in \tabref{tab:results_per_task}, we see lower task load scores for the WYFIWYG conditions. Since the large majority of interactions are due to camera positioning (8122 (cam) vs 99 (rest) avg per participant and task), a lower task load can be linked to better camera controls. 
\begin{table}[tbh]
	\centering
	\setlength{\tabcolsep}{3pt}
	\begin{tabular}[c]{|l||l|l|l|}
		\hline
		Task&Horus&Snapshot&Virt. flight\\
		\hline
		1&37.4$\pm$10.3&\textbf{26.9$\pm$8.3}&36.3$\pm$10.6\\
		2&38.5$\pm$12.2&\textbf{22.2$\pm$6.3}&25.7$\pm$5.9\\
		\hline
	\end{tabular}
	\caption{N.-TLX scores per task with CI-ranges (bold is best).}
	\label{tab:results_per_task}
\end{table}
Interesting to see is the drop-off in task load and the growing result satisfaction over the two tasks for the \textit{virtual-flight} condition, suggesting a steep learning curve for this mode.
The lower task load of WYFIWYG conditions is also supported by lower execution times of T1 in \textit{snapshot} (476.25 sec, SD = 398.74) and \textit{virtual-flight} (584.5 sec, SD = 432.82), compared to \textit{Horus} (669.25 sec, SD = 471.18).

\section{DISCUSSION}\label{sec:discussion}
In this chapter we discuss the findings of work, summarized as implications for the design of UIs and optimization schemes of future quadrotor camera tools. We split the discussion into UI and optimization related aspects. Participant statements come from the exit interview and the thinking aloud protocol.

\subsection{UI Design}
\emph{Visualizing and manipulating the camera path:}
Our general idea of setting the focus on the video content rather than the trajectory was appreciated by our participants with statements like \emph{``in WYFIWYG I think more about what I can do with the camera because I see what it is seeing''}, or \emph{``[...]in WYFIWYG you focus more on the shot''}. One participant also commented positively on the simplicity of WYFIWYG implying that a single view reduces levels of abstraction: \emph{``In Horus you need to abstract more, you need to think where you are in space. With WYFIWYG it's more intuitive''}. 
Nevertheless, 9 out of 12 participants mentioned the need for a 2D-map like in Horus. They highlighted its importance to identify discrepancies of distances in time and space or to specify straight movements in-between keyframes.
Horus' feature of visualizing the camera motion on progress curves caused contradicting reactions.
While some participants perceived them as complicated, others (e.g. the expert) appreciated the workflow they enable, setting camera positions first and then adjust their timing to achieve intended dynamics. 
We propose that future quadrotor camera tools should implement the 3D view as main component of the user interface but also need to provide a 2D map, e.g. as a world-in-miniature rendering (as proposed by participants). In addition, providing progress curves as on-demand feature allows experienced users to manually fine tune camera dynamics while novices are not deterrent by their complexity.

\medskip
\emph{Virtual flight:} Similar accurate results compared to other conditions in T1 and better results in terms of smooth camera motion indicate the value of adjusting target framing continuously. Participants valued the fact that with \emph{virtual flight} they have full control on camera motion: \emph{``In virtual flight I always knew what will happen''}. This positive view was shared by the expert participant: \emph{``Its nice that I can specify movements and that I don't need to think in terms of keyframes and what to do next''}. 
Nevertheless, the high task load scores of this mode in T1 show that practice is necessary in order to use it. 
Therefore, we propose that future quadrotor camera tools should provide \emph{virtual flight} in addition to a keyframe-based camera path specification approach.

\medskip
\emph{Integrated camera control:} We argue that the lower task load values of WYFIWYG conditions compared to Horus are mainly caused by the difference in virtual camera control. In addition, we assume that the better exploratory behavior of WYFIWYG conditions is largely due to the integrated camera control as it gamifies interaction. This was also perceived by participants who commented on using WYFIWYG with \emph{``feels like a game''} or \emph{``is like playing a video game''}.
Therefore, we propose that future quadrotor camera tools should provide integrated positional and rotational camera control.

\subsection{Optimization Scheme Design}
\emph{Target framing:} Undesired camera tilt and non-smooth camera motion due to generating the camera orientation based on look-at positions (referring to section Method, Target Framing) became a problem for two participants. Therefore, we suggest that quadrotor camera tools optimize camera orientations based on reference angles instead of look-at positions.  

\medskip
\emph{Global smoothness:} Existing methods do not optimize the timing of keyframes causing users difficulties in specifying smooth camera motion over an entire sequence. Our observations indicate that most participants did not think about keyframes in space and time, but expected the underlying method to automatically generate globally smooth camera motion over all specified spatial positions. 
The method proposed in this paper somewhat achieves this goal but long optimization runtimes prevented adaption.
We think that reformulating the quadrotor camera trajectory optimization problem to automatically generate timings such that the camera moves smoothly through all user-specified positions would be a more user-friendly approach. 
This could be implemented by optimizing progress on a time-free trajectory subject to a quadrotor's model, similar to \cite{Nageli:2017:SIGGRAPH}. 
Please note that this does not conflict with the requirement of giving users precise timing control, established in \cite{Joubert2015}. 
The suggested workflow is to produce a feasible trajectory with generated timings. These timings should then be editable via progress curves or other means, with a second optimization method guaranteeing that the trajectory remains feasible or returning the closest feasible match (cf. \cite{roberts:2016}).
Investigating the potential of such a method poses an interesting direction for future work.


\section{CONCLUSION}\label{sec:conclusion}
In this paper we investigate how to improve end-user support in quadrotor camera tools. We highlight important aspects for the creation of aesthetically pleasing aerial footage, revealed in formative expert interviews. Based on these results, we design a new quadrotor camera tool, WYFIWYG, and develop extensions to an existing trajectory generation algorithm that allow for the generation of more intuitive angular camera motion and globally smooth trajectories over a sequence of keyframes. To better understand the effectiveness of particular UI- and optimization scheme features in terms of user support, we conduct an exploratory user study evaluating variants of our system and \cite{Joubert2015}. The study revealed that current tools complicate the design of globally smooth video shots by requiring users to specify keyframes at equidistant points in time and space. We conclude by discussing the findings of work and summarizing them as implications for the design of UIs and optimization schemes of future quadrotor camera tools.

\section{ACKNOWLEDGEMENTS}\label{sec:conclusion}
This work was funded in parts by the Swiss National Science Foundation (UFO 200021L\_153644).

\section{Appendix A - Approximate Quadrotor Model and Trajectory Generation}
For algorithmic motion plan generation a model of the quadrotor and its dynamics are needed. Incorporating a fully non-linear model results in a high computational cost and negates convergence guarantees \cite{Mellinger2011a}. Following \cite{Gebhardt:2016:Airways} we use a  linear approximation, modelling the quadrotor as a rigid body, described by its mass and moment of inertia along the world frame z-axis (i.e.\ pitch and roll are fixed):
\begin{align}
	\label{eq:quad_dynamical_system_continuous}
	m\ddot{\mathbf{r}} = \mathbf{F} + m \mathbf{g} & \in \mathbb{R}^3 \\
	I_{\psi} \ddot{\psi_q} = M_{\psi} & \in \mathbb{R} , \nonumber
\end{align}
where $\mathbf{r}$ is the center of mass,
$\psi_q$ is the yaw angle, $m$ is the mass of the quadrotor,
$I_{\psi}$ is the moment of inertia about the z-axis,
$\mathbf{u}_{r}$ is the the force acting on $\mathbf{r}$
and $M_{\psi}$ is the torque along $z$.

To ensure that robot and gimbal can reach specified positions and camera orientations within a given time and without exceeding the limits of the quadrotor hardware, bounds on maximum force and torque are introduced:
\begin{equation}
\mathbf{u}_{\mathit{min}} \leq \mathbf{u} \leq \mathbf{u}_{\mathit{max}} \in \mathbb{R}^4  ,
\label{eq:input_bounds}
\end{equation}
where $\mathbf{u} = [\mathbf{F}, M_{\psi}]^{T}$ is the input to the system. Details on how to choose  the linear bounds can be found in~\cite{Gebhardt:2016:Airways}.
This quadrotor model is reformulated as a first-order dynamical system and discretized in time with a time-step $\Delta t$ assuming a zero-order hold strategy, i.e. keeping inputs constant in between stages:
\begin{equation}
	\label{eq:quad_dynamical_system_discrete}
	\mathbf{x}_{i+1} = A_d \mathbf{x}_{i} + B_d \mathbf{u}_{i} + c_{d} ,
\end{equation}
where $\mathbf{x}_{i} = [\mathbf{r}, \psi, \dot{\mathbf{r}}, \dot{\psi}]^{T} \in \mathbb{R}^{4}$
is the state and $\mathbf{u}_{i}$ is the input of the system at time $i \Delta t$. The matrix
$A_d \in \mathbb{R}^{8x8}$ propagates
the state $\mathbf{x}$ forward by one time-step, the matrix
$B_d \in \mathbb{R}^{8x4}$ describes the effect of the
input $\mathbf{u}$ on the state and the vector
$c_{d} \in \mathbb{R}^{8}$ that of gravity after one time-step.

The algorithm takes $M$ positions $k_{j}$
at a specific time $\eta(j) \Delta t$ as input, where $\eta: \mathbb{N} \rightarrow \mathbb{N}$ maps between keyframe indices and corresponding time-point. Time is discretized into $N$ stages with stepsize $\Delta t$ over the whole time horizon $[0, t_{f}]$.
The variables which are optimized are the quadrotor state $x_{i}$ and the inputs $u_{i}$ to the system \eqref{eq:quad_dynamical_system_discrete} at each stage $i \Delta t$.
For the camera motion to follow the user-specified positions as closely as possible, we seek to minimize the following cost
\begin{equation}
\label{eq:keyframe_cost}
E^{k} = \sum_{j=1}^{M} || r_{\eta(j)} - k_{j} || ^ {2} .
\end{equation}
A small residual of $E^{k}$ indicates a good match of the generated quadrotor position and the specified
keyframe.
Furthermore, we wish to generate smooth motion, which is related to the derivatives of the quadrotor's position. To this end we introduce a cost for penalizing higher position derivatives
\begin{equation}
\label{eq:quad_derivative_cost}
E^{d} = \sum_{i=q}^{N} || D^{q} \begin{bmatrix}
x_{i} \\
\ldots \\
x_{i-q} \\
\end{bmatrix} || ^ {2} ,
\end{equation}
where $D^{q}$ is a finite-difference approximation of the $q$-th derivative over the last $q$ states.
The combined cost $E = \lambda_{k} E^{k} + \lambda_{d} E^{d}$ with weights $\lambda_{k|d}$ is
a quadratic function, enabling us to formulate the trajectory generation problem as a quadratic program.
\begin{align}
	\label{eq:quadratic_program}
	\underset{X}{\text{minimize}} \ & \frac{1}{2} X^{T} H X + f^{T} X  \\
	\text{ subject to } & A_{\mathit{ineq}} X \leq b_{\mathit{ineq}} \nonumber \\
	\text{ and } & A_{\mathit{eq}} X = b_{\mathit{eq}} \nonumber ,
\end{align}
where $X$ denotes the stacked state vectors $x_{i}$ and inputs $u_{i}$ for each time-point, $H$ and $f$ contain the
quadratic and linear cost coefficients respectively which are defined by \eqref{eq:keyframe_cost} and \eqref{eq:quad_derivative_cost}
, $A_{\mathit{ineq}}$, $b_{\mathit{ineq}}$ comprise the linear inequality
constraints of the inputs \eqref{eq:input_bounds} and
$A_{\mathit{eq}}$, $b_{\mathit{eq}}$ are the linear equality constraints
from our model \eqref{eq:quad_dynamical_system_discrete}
for each time-point $i \in {1,\ldots,N}$.
This problem has a sparse structure and can be solved by most optimization software packages.

\section{Appendix B - UEQ Scores and Tool Preference}
We also asked participants to fill out the User Experience Questionnaire (UEQ). Its scores reveal a distinct ranking in between conditions. \emph{Snapshot} ranks first on all dimensions, followed by \emph{virtual flight} and \emph{Horus} (see \tabref{tab:ueq}). Reasoning about the cause of the scores is difficult. We assume that the higher level of attractiveness of the WYFIWYG-conditions is caused by the simplicity of the UI, having a single view to design the video. The better efficiency scores of the WYFIWYG-conditions are likely caused by the integrated camera control.
Finally, we asked participants which condition they prefer. 9 out of 12 participants preferred WYFIWYG ($6\times$\textit{snapshot}, $2\times$\textit{virtual-flight}, $1\times$either) with the remaining 3 stating equal preference for Horus and one of the WYFIWYG conditions.
\begin{table}[tbh]
	\centering
	\setlength{\tabcolsep}{3pt}
	\begin{tabular}[c]{|l||l|l|l|}
		\hline
		Dimension&Horus&Snapshot&Virtual flight\\
		\hline
		Attractiveness&0.35$\pm$0.64&\textbf{1.91$\pm$0.39}&1.19$\pm$0.72\\
		Perspicuity&-0.29$\pm$0.61&\textbf{2.0$\pm$0.32}&1.48$\pm$0.59\\
		Efficiency&0.42$\pm$0.54&\textbf{1.52$\pm$0.41}&1.13$\pm$0.59\\
		Dependability&0.56$\pm$0.54&\textbf{1.38$\pm$0.43}&0.52$\pm$0.59\\
		Stimulation&0.73$\pm$0.49&\textbf{1.63$\pm$0.5}&1.4$\pm$0.53\\
		Novelty&0.25$\pm$0.86&\textbf{1.31$\pm$0.62}&1.13$\pm$0.56\\
		\hline
	\end{tabular}
	\caption{UEQ dimension scores with CI-ranges (bold is best).}
	\label{tab:ueq}
\end{table}

\bibliographystyle{SIGCHI-Reference-Format}
\bibliography{bib/references}

\balance{}

\end{document}